\documentclass[conference]{IEEEtran}

\IEEEoverridecommandlockouts

\usepackage{graphicx}
\usepackage{amsmath}
\usepackage{amssymb}
\usepackage{color}

\usepackage{algorithm}
\usepackage{algpseudocode}
\algnewcommand\algorithmicinput{\textbf{Input:}}
\algnewcommand\INPUT{\item[\algorithmicinput]}
\algnewcommand\algorithmicoutput{\textbf{Output:}}
\algnewcommand\OUTPUT{\item[\algorithmicoutput]}

\usepackage{mathtools}

\newcommand{\Fig}[1]{Fig.~\textup{\ref{#1}}}
\graphicspath{{figures/}}
\usepackage{subcaption}

\def\BibTeX{{\rm B\kern-.05em{\sc i\kern-.025em b}\kern-.08em
    T\kern-.1667em\lower.7ex\hbox{E}\kern-.125emX}}

\usepackage[left=0.625in,right=0.625in,
    top=0.75in,bottom=1.0in,bindingoffset=0cm]{geometry}
\begin{document}

\title{ High Performance Interference Suppression \\
in Multi-User Massive MIMO Detector \thanks{The research was carried out at Skoltech and supported by the Russian Science Foundation (project no. 18-19-00673).}}

\author{
\IEEEauthorblockN{Andrey Ivanov, Alexander Osinsky, Dmitry Lakontsev, Dmitry Yarotsky}
\IEEEauthorblockA{\small Skolkovo Institute of Science and Technology\\
    Moscow, Russia
}
  { an.ivanov@skoltech.ru, Alexander.Osinsky@skoltech.ru, d.lakontsev@skoltech.ru, D.Yarotsky@skoltech.ru }  
  {} 
}


%


\maketitle

\begin{abstract}

In this paper, we propose a new nonlinear detector with improved interference suppression in Multi-User Multiple Input, Multiple Output (MU-MIMO) system. The proposed detector is a combination of the following parts: QR decomposition (QRD), low complexity users sorting before QRD, sorting-reduced (SR) K-best method and minimum mean square error (MMSE) pre-processing. Our method outperforms a linear interference rejection combining (IRC, i.e. MMSE naturally) method significantly in both strong interference and additive white noise scenarios with both ideal and real channel estimations. This result has wide application importance for scenarios with strong interference, i.e. when co-located users utilize the internet in stadium, highway, shopping center, etc. Simulation results are presented for the non-line of sight 3D-UMa model of 5G QuaDRiGa 2.0 channel for 16 highly correlated single-antenna users with QAM16 modulation in 64 antennas of Massive MIMO system. The performance was compared with MMSE and other detection approaches. 

\end{abstract}

 \vskip 0.2cm


\section{Introduction}

The fifth generation (5G) of wireless systems will demand more users with a much higher overall capacity \cite{A1A}. In recent years, multi-user Massive Multiple Input, Multiple Output (MU-MIMO) and massive MIMO have been adopted as the key technologies to address the capacity requirements of enhanced Mobile Broadband (eMBB) in $5$G \cite{A1A, A3}. MIMO detection \cite{A2} is a method of antennas digital signal processing to extract user signals in an uplink channel of a base station. Compared to conventional MIMO systems, which have already reached their throughput limits, massive MIMO has become the most promising candidates to increase transmission data rate over wireless networks. In massive MIMO systems, each base station is equipped with tens or hundreds of antennas dedicated to serving tens of users. It has been proved \cite{A3} that a massive MIMO system can increase the spectrum efficiency of a wireless channel by several times. Spatially multiplexed MIMO systems can support several independent data streams, resulting in a significant increase of the system throughput \cite{A1A, A3} due to multiple spectrum reuse. In this context, a great deal of effort has been made in the development of multi-user Massive MIMO detection method, which is robust to unknown interference from users of other (neighbour) cells \cite{A20}. 

\subsection{ML detector}

The maximum likelihood (ML) detector performs an exhaustive search by calculating the Euclidean distance for every possible symbol vector candidate. The number of candidate symbol vectors grows exponentially with the number of users and the number of bits per constellation point \cite{A1,A45,A46}. Thus, with high order constellations (QAM$64$ or higher) in MU-MIMO mode, ML detection becomes computationally heavy. The ML estimation is given in the frequency domain by \cite{A2}:
\begin{equation} \label{GrindEQ__1_} 
\boldsymbol{y}=\boldsymbol{H}\boldsymbol{x}+\boldsymbol{n}+\boldsymbol{i},  
\end{equation} 
\[\boldsymbol{x}_{ML} =\underset{x} \arg \min \left\| \boldsymbol{y}-\boldsymbol{H}\boldsymbol{x}\right\| ,\] 
where $\boldsymbol{y}$ is the frequency domain received signal vector of size $N$; $N$ is the number of RX antennas; $\boldsymbol{H}$ is the channel matrix of size $\left[N\times M\right]$, $\boldsymbol{x}$ is the TX vector of size $M$; $M$ is the number of single-antenna users in the uplink channel; $\boldsymbol{n}$ is the white noise; $\boldsymbol{i}$ is the interference (unknown); $\boldsymbol{x}_{ML}$ is the ML estimation of vector $\boldsymbol{x}$. 

\subsection{Single user MMSE detector}

Linear detection methods \cite{A2}, \cite{A44} consider the input-output relation of a MIMO system as an unconstrained linear estimation problem, which can be solved by using minimum mean square error (MMSE) method. The resulting unconstrained estimate ignores the fact that the transmitted symbols are from a limited set of constellation points. Let us describe the baseline MMSE detector for one single antenna user and $N=3$ receiving antennas of the base station. Define $\boldsymbol{h}$ as a vector of frequency domain channel estimations for one subcarrier:
\[\boldsymbol{h}\boldsymbol{=}{\left[ \begin{array}{ccc}
{\hat{h}}_1 & {\hat{h}}_2 & {\hat{h}}_3 \end{array}
\right]}^{\boldsymbol{T}}\] 
Define (interference + noise) signal as follows:
\[{{{u}_k\boldsymbol{=}n}_k+i}_k,\]
where $k$ is the antenna index. Define (interference + noise) covariance matrix as:
\begin{equation} \label{GrindEQ__5_} 
{\boldsymbol{R}}_{\boldsymbol{uu}}=\mathbb{E}\left(\left[ \begin{array}{c}
u_1 \\ 
u_2 \\ 
u_3 \end{array}
\right]\left[ \begin{array}{ccc}
u^*_1 & u^*_2 & u^*_3 \end{array}
\right]\right) 
\end{equation} 
\begin{equation} \label{GrindEQ__6_} 
{\boldsymbol{\mathrm{R}}}_{\boldsymbol{uu}} =\left[ \begin{array}{ccc}
{\mathrm{R}}_{\mathrm{11}} & {\mathrm{R}}_{\mathrm{12}} & {\mathrm{R}}_{\mathrm{13}} \\ 
{\mathrm{R}}_{\mathrm{21}} & {\mathrm{R}}_{\mathrm{22}} & {\mathrm{R}}_{\mathrm{23}} \\ 
{\mathrm{R}}_{\mathrm{31}} & {\mathrm{R}}_{\mathrm{32}} & {\mathrm{R}}_{\mathrm{33}} \end{array}
\right]  
\end{equation} 
Assume interference + noise matrix ${\boldsymbol{R}}_{\boldsymbol{uu}}$ has the same value inside each resource block. This assumption is quite valid since the interference power from other cells is also approximately the same inside each resource block in frequency domain and the same inside time transmission interval in the time domain. Therefore, in case of smooth channel response inside the mentioned time and frequency unit, element of ${\boldsymbol{R}}_{\boldsymbol{uu}}$ matrix in equation \eqref{GrindEQ__5_} can be estimated as:
\[{\widehat{{R}}_{uu}}(j,k)=\mathbb{E}(u_ju_k^*) ,\] 
where $j,k$ are antenna indexes; $u^*$ is the complex conjugate of $u$. In practice, $u_k$ can be estimated with reference signals as:
\[{\widehat{u}}_k={y}_k-{\hat{h}}_kx\] 
The MMSE detection algorithm is intended to minimize $e^2={\left|x-\boldsymbol{\ }\widehat{x}\right|}^{2}$ and can be obtained as a frequency domain Wiener filter \cite{A5}, as expressed by the classical equations:
\[\widehat{x}={\boldsymbol{w}}\boldsymbol{y}\]
\begin{equation} \label{GrindEQ__7_} 
\boldsymbol{w}={\boldsymbol{R}}^{\boldsymbol{-}\boldsymbol{1}}_{\boldsymbol{yy}}\boldsymbol{h^{\boldsymbol{H}},} 
\end{equation} 
where $\boldsymbol{w}$ is the weight vector, $\boldsymbol{h}$ is the channel estimation vector as described in \cite{A50} and \cite{A51}, $\boldsymbol{y}\boldsymbol{=}{\left[ \begin{array}{ccc}
{y}_1 & {y}_2 & {y}_3 \end{array}
\right]}^{\boldsymbol{T}}$, covariance matrix ${\boldsymbol{R}}_{\boldsymbol{yy}}$ of the received signal is defined similarly to \eqref{GrindEQ__5_}. The matrix ${\boldsymbol{R}}_{\boldsymbol{yy}}$ can also be calculated as:
\begin{equation} \label{GrindEQ__8_} 
{\boldsymbol{R}}_{\boldsymbol{yy}}={\boldsymbol{R}}_{\boldsymbol{uu}}+\boldsymbol{h}{\boldsymbol{h}}^{\boldsymbol{H}} 
\end{equation} 
From equations \eqref{GrindEQ__7_} and \eqref{GrindEQ__8_} using the Sherman--Morrison formula we can derive the MMSE estimation as:
\begin{equation} \label{GrindEQ__9_} 
\boldsymbol{w}=\frac{{\boldsymbol{h}}^{\boldsymbol{H}}{\boldsymbol{R}}^{\boldsymbol{-1}}_{\boldsymbol{uu}}}{1+{\boldsymbol{h}}^{\boldsymbol{H}}{\boldsymbol{R}}^{\boldsymbol{-}\boldsymbol{1}}_{\boldsymbol{uu}}\boldsymbol{h}} 
\end{equation} 
In equation \eqref{GrindEQ__8_} the matrix ${\boldsymbol{R}}_{\boldsymbol{uu}}$ is an interference plus noise matrix, which is responsible for interference suppression, i.e. interference rejection combining (IRC). It focuses null of the $\boldsymbol{w}$ pattern to the interference sources to suppress it, while $\boldsymbol{h}$ focuses the main beam to the user, i.e. Maximum Ratio Combining (MRC).

\subsection{Multi-user MMSE detector}

Assume we have $M$ single antenna users per subcarrier. In this case the MMSE detector is given by:
\[\boldsymbol{\widehat{x}} =\boldsymbol{W}\boldsymbol{y}, \] 
\begin{equation}\label{GrindEQ__91_} 
\boldsymbol{W}={(\sigma_{n}^{2} \boldsymbol{I} + \boldsymbol{H}^{H}{\boldsymbol{R}}^{\boldsymbol{-1}}_{\boldsymbol{uu}}\boldsymbol{H}) }^{-1}{\boldsymbol{H}^{\boldsymbol{H}}{\boldsymbol{R}}^{\boldsymbol{-1}}_{\boldsymbol{uu}} }, 
\end{equation}
where $\boldsymbol{\widehat{x}}$ is the linear estimation of the frequency domain vector $\boldsymbol{x}$; $\boldsymbol{W}$ is the weight matrix of size $\left[N\times M\right]$; $\sigma_{n}^{2}$ is the RX antennas noise power; $\boldsymbol{I}$ is the identity matrix of size $\left[M\times M\right]$. In case of $\boldsymbol{i} = \boldsymbol{0}$, equation \eqref{GrindEQ__91_} represents the MRC detector:
\begin{equation}\label{GrindEQ__92_} 
\boldsymbol{W}={(\sigma_{n}^{2} \boldsymbol{I} + \boldsymbol{H}^{\boldsymbol{H}}\boldsymbol{H})}^{-1} {\boldsymbol{H}^{\boldsymbol{H}} }
\end{equation}
Linear detection schemes are simple, but unfortunately, they do not consider the lattice structure of the transmitted complex amplitudes $\boldsymbol{x}$, and, therefore, do not provide good enough performance, especially when the channel matrix is near singular.

\subsection{MMSE OSIC detector}

MMSE with ordered successive interference cancellation (MMSE-OSIC) detection is performed with QR decomposition (QRD) of the permuted channel matrix $\boldsymbol{H}_{perm} $, which is defined as in \cite{A5}, \cite{A6} and \cite{A7}:

\[\boldsymbol{y}_{ext} =\boldsymbol{H}_{perm} \boldsymbol{x}_{perm} +\boldsymbol{n}+\boldsymbol{i}, \] 
\[\boldsymbol{x}_{perm} =\boldsymbol{P}\boldsymbol{x}, \] 
\begin{equation}\label{eq10}
\boldsymbol{H}_{perm} =\boldsymbol{P}\boldsymbol{H}_{ext},   
\end{equation}
\[\boldsymbol{H}_{ext} =\left[\begin{array}{cc} {\boldsymbol{H}^{\boldsymbol{T}} } & {\sqrt{\sigma_{n}^2+\sigma_{i}^2} \boldsymbol{I}} \end{array}\right]^{\boldsymbol{T}} ,\]   
\[\boldsymbol{y}_{ext} =\left[\begin{array}{cc} {\boldsymbol{y}^{\boldsymbol{T}} } & {\boldsymbol{0}} \end{array}\right]^{\boldsymbol{T}} , \] 
where $\sigma_{i}^{2}$ is the RX antennas interference power, $\boldsymbol{P}$ is the permutation matrix, $\boldsymbol{x}_{perm}$ is the permuted version of $\boldsymbol{x}$, and $\boldsymbol{H}_{ext}$ matrix is utilized instead of $\boldsymbol{H}$ for the regularization reason. MMSE with ordered successive interference cancellation (OSIC) detector is based on QR factorization of the channel matrix as shown in \cite{A6}:
\[\boldsymbol{H}_{perm} =\boldsymbol{Q}\boldsymbol{R}, \] 
\[\boldsymbol{Q}^{\boldsymbol{H}} \boldsymbol{y}_{ext} =\left(\boldsymbol{Q}^{\boldsymbol{H}} \boldsymbol{Q}\right)\boldsymbol{R}\boldsymbol{x}_{perm}+\boldsymbol{Q}^{\boldsymbol{H}} (\boldsymbol{n}+\boldsymbol{i}) , \] 
\begin{equation} \label{GrindEQ__10_} 
\boldsymbol{Q}^{\boldsymbol{H}} \boldsymbol{y}_{ext} =\boldsymbol{R}\boldsymbol{x}_{perm}+\boldsymbol{Q}^{\boldsymbol{H}} (\boldsymbol{n}+\boldsymbol{i}) , 
\end{equation}
where $\boldsymbol{R}$ is the $\left[\left(N+M\right)\times M\right]$ upper triangular matrix; $\boldsymbol{P}$ is the permutation matrix. Detection starts with $\boldsymbol{x}_{perm} \left(M\right)$ amplitude detection and stops after $\boldsymbol{x}_{perm} \left(1\right)$ calculation according to the upper triangle matrix $\boldsymbol{R}$ structure. Therefore, the initial vector $\boldsymbol{x}$ can be derived as $\boldsymbol{x}=\boldsymbol{P}^{T} \boldsymbol{x}_{perm}$. The MMSE-OSIC method demonstrates better performance in comparison with the MMSE detection, but the gain is limited due to error propagation, caused by non-ideal user sorting before QRD, high correlation among layers and imperfect channel estimation. The users sorting is intended to reorder diagonal elements of the upper triangular matrix $\boldsymbol{R}$ in ascending order to prevent error propagation in MU-MIMO scenario.

Using only MMSE-OSIC doesn't achieve the best performance in multi-user scenarios of $16 \times 64$ MIMO system. Therefore, a post-processing K-best algorithm should be used to enhance performance in acceptable complexity, as described in \cite{A1, A45, A46}. Sorting reduced K-best (SR-K-best in \cite{A1}) is a version of the K-best with low sorting complexity. Sorting the best $K$ survivors from $KM$ candidates, where $M=4$, is reduced to sorting the best $S$ from much less number of candidates, while the residual $K-S$ survivors are defined as "most expected" and taken from the full candidates set of size $KM$ before the sorting according to a special selection algorithm. The paper \cite{A1} proposes SR-K-best with parameters $(K,S,\boldsymbol{p})$. The vector $\boldsymbol{p}$ means the positions of the "most expected" candidates. Unfortunately, the SR-K-best with $(K, S,\boldsymbol{p})$ parameters also results in performance losses in high correlated scenarios. Therefore, we utilize a new flexible structure $(K,S,\boldsymbol{p},\boldsymbol{v},\boldsymbol{q})$ of SR-K-best algorithm \cite{A45, A46}. The vector $\boldsymbol{v}$ defines a set of sorting child nodes (i.e. a set for $S$ sorted candidates search); while the vector $\boldsymbol{q}$ defines the location of $S$ sorted candidates in the final composed list of $K$ candidates for the next detection iteration.

Finally, MMSE-OSIC with the optimized user sorting and the SR-K-best demonstrates performance close to the maximum likelihood algorithm in the additive white noise channel, i.e. in case of $\boldsymbol{i}=\boldsymbol{0}$ in \eqref{GrindEQ__10_}. However, the mentioned MMSE-OSIC is quite sensitive to external interference from unknown users (when $\boldsymbol{i} \neq \boldsymbol{0}$), while the basic MMSE algorithm is robust to the correlated noise due to the IRC algorithm with the matrix ${\boldsymbol{R}}_{\boldsymbol{uu}}$. To overcome the interference suppression problem, we propose a new pre-processing algorithm.   

\section{Simulation tool}

\begin{figure}[h]
\centering
\includegraphics[width=1.0\columnwidth]{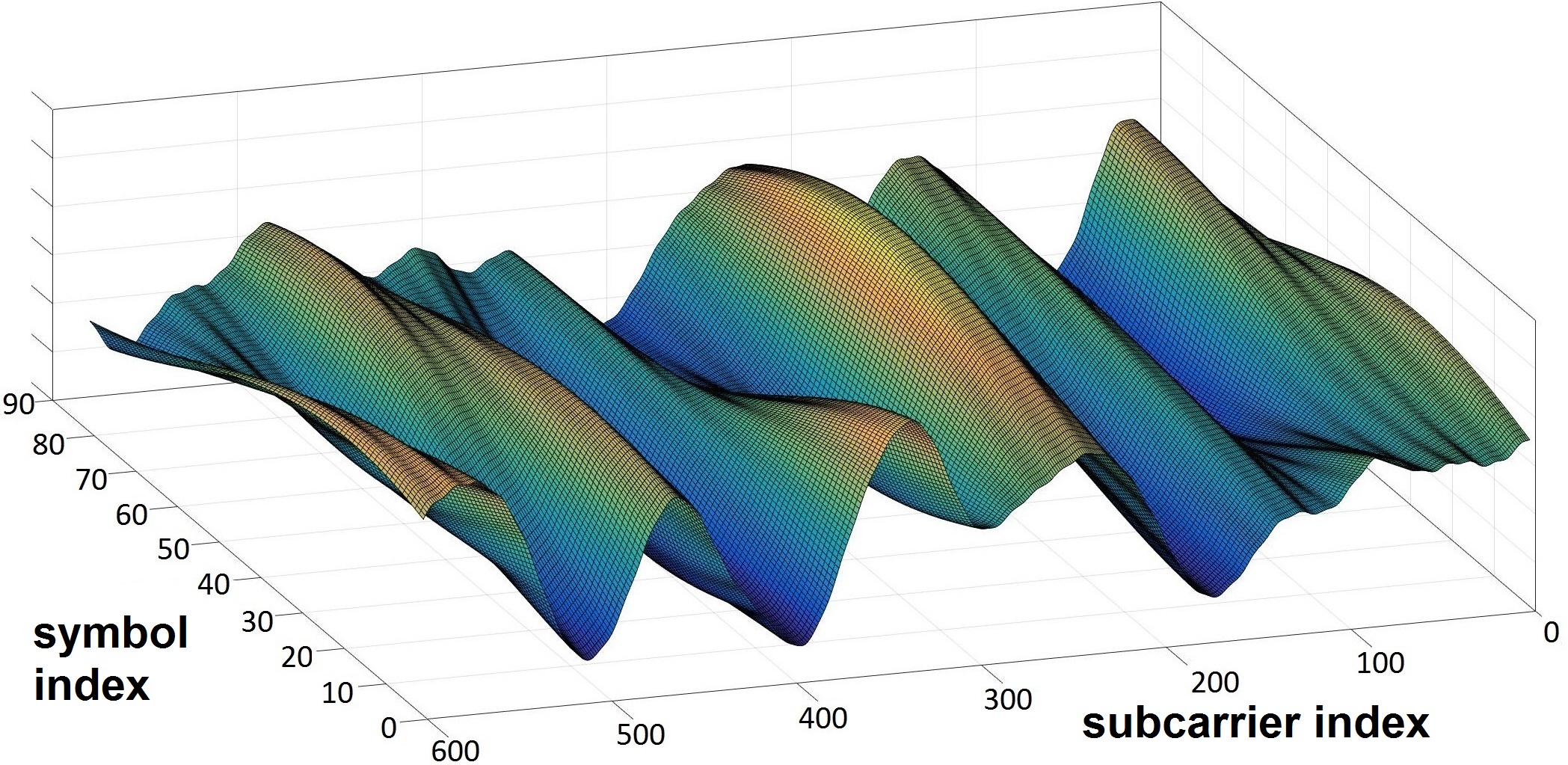}
\caption{
Magnitude spectrum of QuaDRiGa channel
}
\label{fig1}
\end{figure}

QuaDRiGa, short for "QUAsi Deterministic RadIo channel GenerAtor" \cite{A13}, is mainly used to generate realistic radio channel responses for use in system-level simulations of 5G mobile networks. We test our algorithms in high correlated 3D-UMa, BERLIN-UMa and DRESDEN-UMa non-line of sight (NLOS) models of MIMO $16 \times 64$ scenario for users with a speed of $5$ km/h. The number of users = $16$ is chosen as the max number of active users per subcarrier in a cross-polarized $64$ RX antennas system according to $5$G standard \cite{A3}. The antenna array is mounted on $25$ meters high and consists of $2$ co-located rectangle subarrays with $32$ antennas each. The carrier frequency is $3.5$GHz, a maximum distance to the user is $500$ meters. The interference is given by $4$ unknown users with QAM16 modulation and approximately the same power as target users. A short $3D$ fragment of the channel magnitude spectrum is shown in \Fig{fig1}.

\section{SR-K-best detector}

\begin{figure}[h]
\centering
\includegraphics[width=1.0\columnwidth]{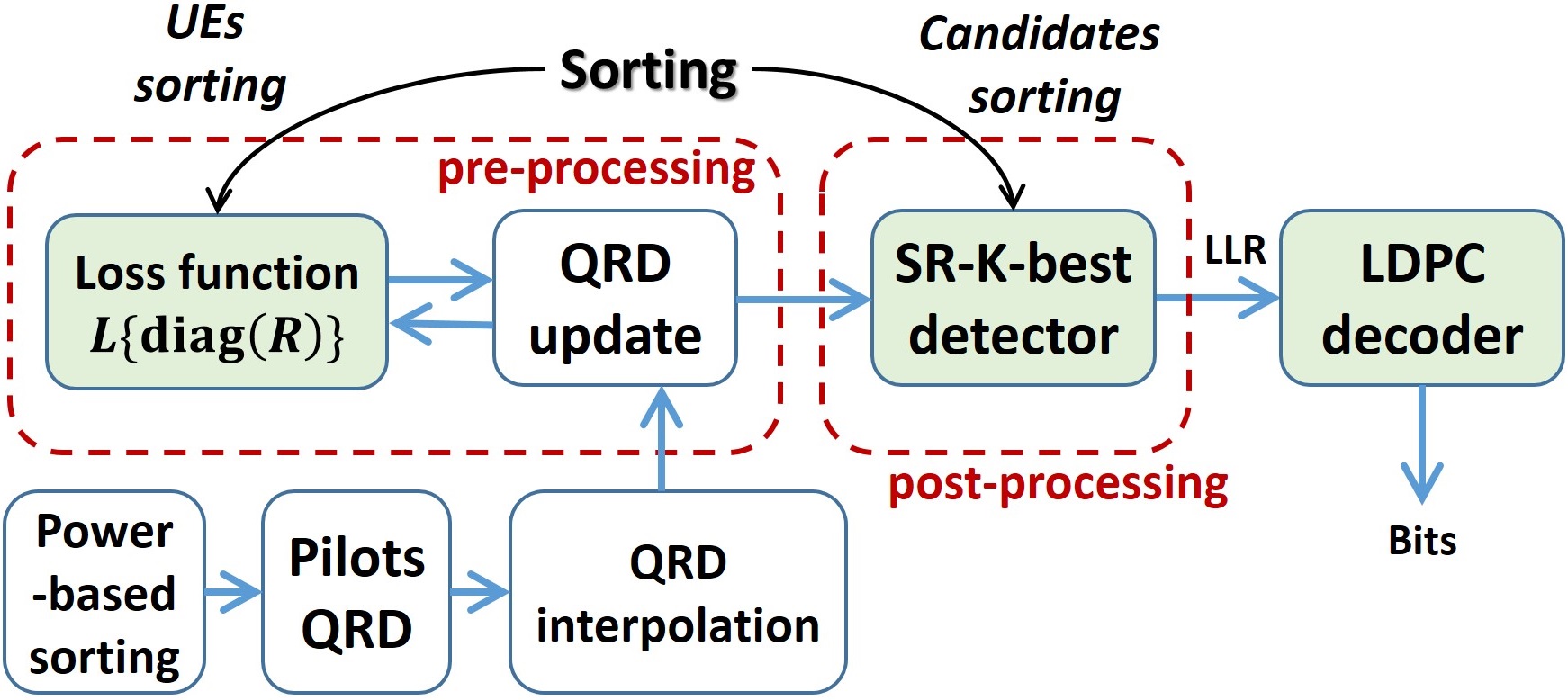}
\caption{
Detector scheme
}
\label{fig2}
\end{figure}

A functional scheme of the QRD-based detection approach is shown in \Fig{fig2}. It consists of pre-processing and post-processing parts \cite{A1, A45}. Pre-processing is required to calculate sorted QRD in two steps. 

$\boldsymbol{Step}$ $\boldsymbol{1}$: QRD interpolation, as described in \cite{A14}.

For QRD calculation in the MIMO system, we have to perform the QRD for each subcarrier. In practice, the interpolation-based QRD only computes the $\boldsymbol{Q}$ and $\boldsymbol{R}$ matrixes for the pilot subcarriers to reduce computational complexity. Then, the $\boldsymbol{Q}$ and $\boldsymbol{R}$ of the data subcarriers are interpolated from those of the pilot subcarriers. 

$\boldsymbol{Step}$ $\boldsymbol{2}$: users sorting (strings permutation in $\boldsymbol{H}_{ext}$) is required to achieve the permutation matrix $\boldsymbol{P}$ in equation \eqref{eq10}.

Users sorting problem is well-known, for example, a post-sorting algorithm and pre-sorting solution are analyzed in \cite{A6}. We take $\boldsymbol{P}$ matrix from QRD of the pilot symbols as the first step of $\boldsymbol{P}$ matrix calculation for the data symbol to realize sorting track as proposed in \cite{A45}. The $\boldsymbol{P}$ matrix changes very slowly from one symbol to another, and a low sorting complexity is required to update it. Loss function $L=L\{\mathrm{diag}\left(\boldsymbol{R}\right)\}$ of diagonal entries of interpolated $\boldsymbol{R}$ matrix is used to guarantee the least number of matrix $\boldsymbol{P}$ updates.  

Post-Processing part is represented by SR-K-best detection. The $(K,S,\boldsymbol{p},\boldsymbol{v},\boldsymbol{q})$ parameters were optimized in \cite{A45} for QuaDRiGa channel of $16 \mathrm{\times} 64$ MIMO system and given by:
\begin{multline}  \label{all}
(K,S)=(16,4), \\
\boldsymbol{p}\mathrm{=[2\ 1\ 1\ 1\ 1\ 1\ 1\ 1\ 1\ 1\ 1\ 0\ 0\ 0\ 0\ 0]},\\
\boldsymbol{v}\mathrm{=[2\ 1\ 1\ 1\ 1\ 1\ 1\ 1\ 1\ 1\ 1\ 2\ 2\ 2\ 2\ 2]},\\
\boldsymbol{q}=\left[2\ 4\ 6\ 8\right].
\end{multline}

\begin{figure}[h]
\centering
\includegraphics[width=0.9\columnwidth]{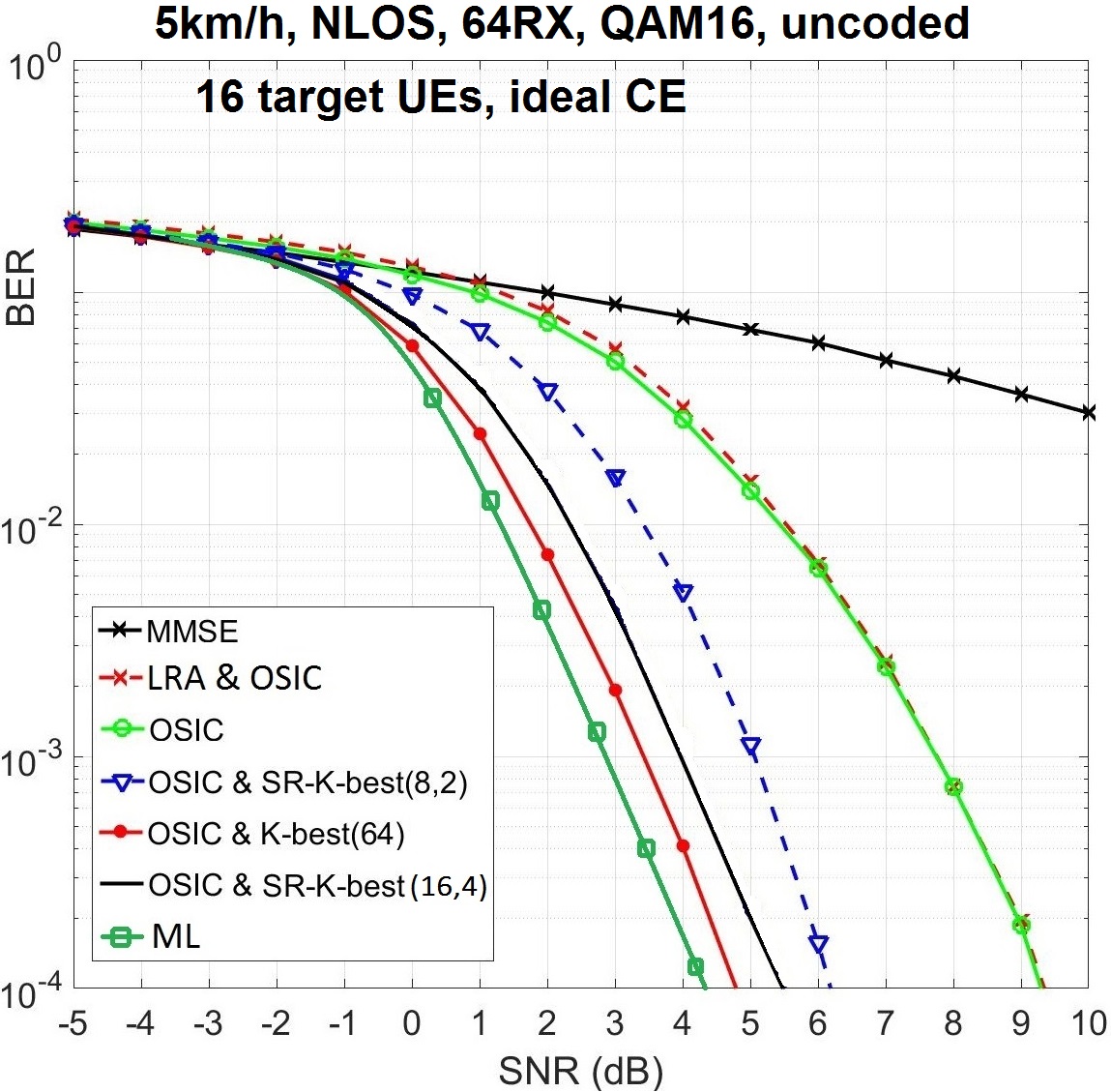}
\caption{
Uncoded BER in AWGN channel with ideal CE
}
\label{fig2_1}
\end{figure}

\begin{figure}[h]
\centering
\includegraphics[width=0.9\columnwidth]{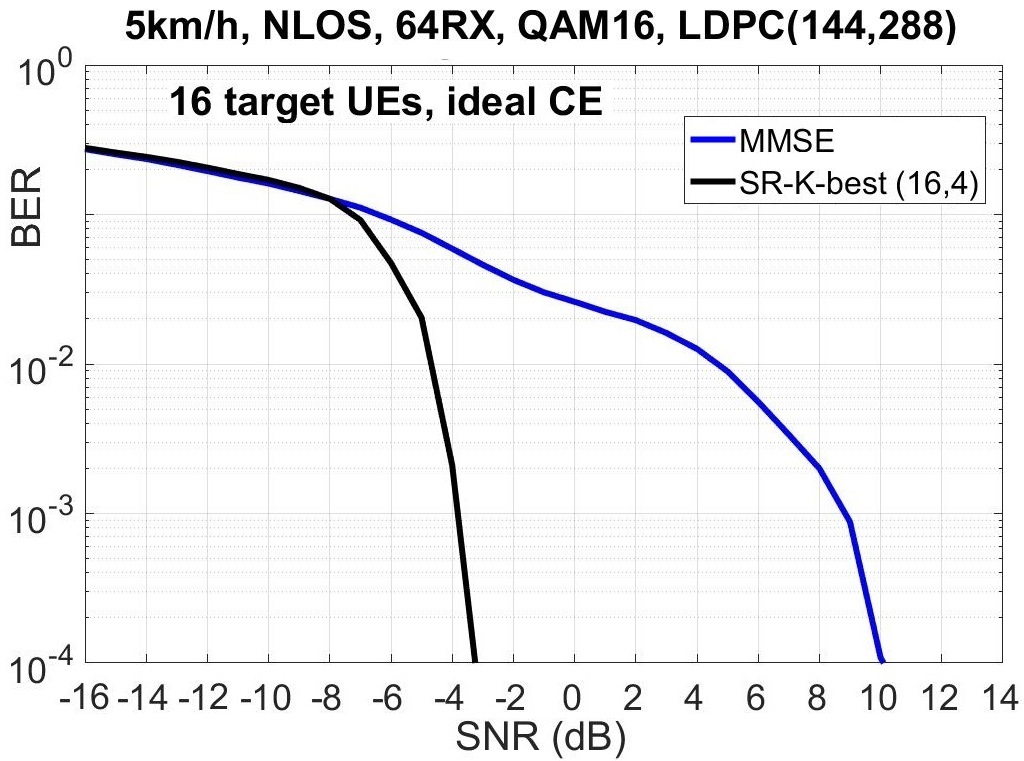}
\caption{
Coded BER in AWGN channel with ideal CE
}
\label{fig3}
\end{figure}

\begin{figure}[h]
\centering
\includegraphics[width=0.9\columnwidth]{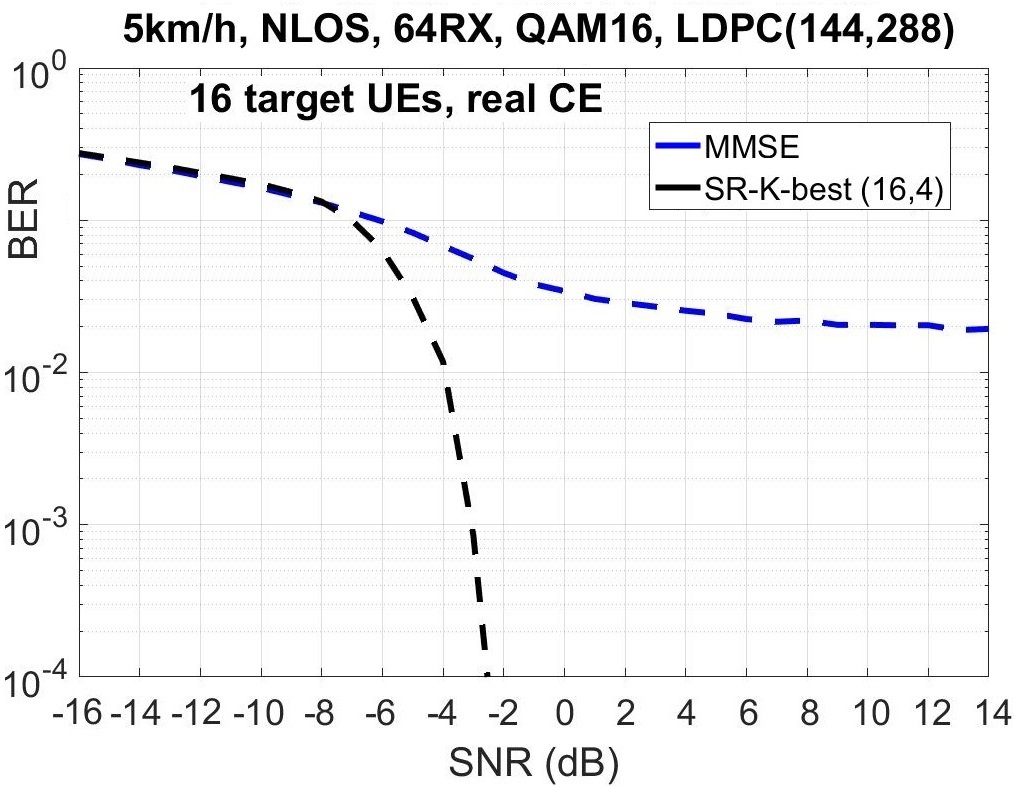}
\caption{
Coded BER in AWGN channel with real CE
}
\label{fig4}
\end{figure}

Simulation results for Lattice Reduction Aided (LRA) OSIC detector \cite{A1}, MMSE, ML and some other state-of-the-art detecton algorithms are presented in additive white noise (AWGN) channel in \Fig{fig2_1} for $16$ target users, bit error rate (BER) is given for uncoded case \cite{A45}. The SR-K-best detector of type $(K,S,\boldsymbol{p},\boldsymbol{v},\boldsymbol{q})$ with $(K,S)=(16,4)$ parameters demonstrates performance, close to ML. In \Fig{fig3} and \Fig{fig4} simulation results are presented for the same $16$ coded users in scenario with low-density parity-check (LDPC) decoder with $(144,288)$ code in the MIMO receiver for ideal and real channel estimations (DFT-based channel estimation from \cite{A50} was implemented). The min-sum decoding algorithm was utilized as described in \cite{A16}. The SR-K-best detector outperforms the MMSE detector in the AWGN channel for both ideal and real channel estimation. Therefore, the SR-K-best detector with $(K,S)=(16,4)$ parameters is a good choice for the 5G receiver.

\section{Robust SR-K-best detector}

\begin{figure}[h]
\centering
\includegraphics[width=0.9\columnwidth]{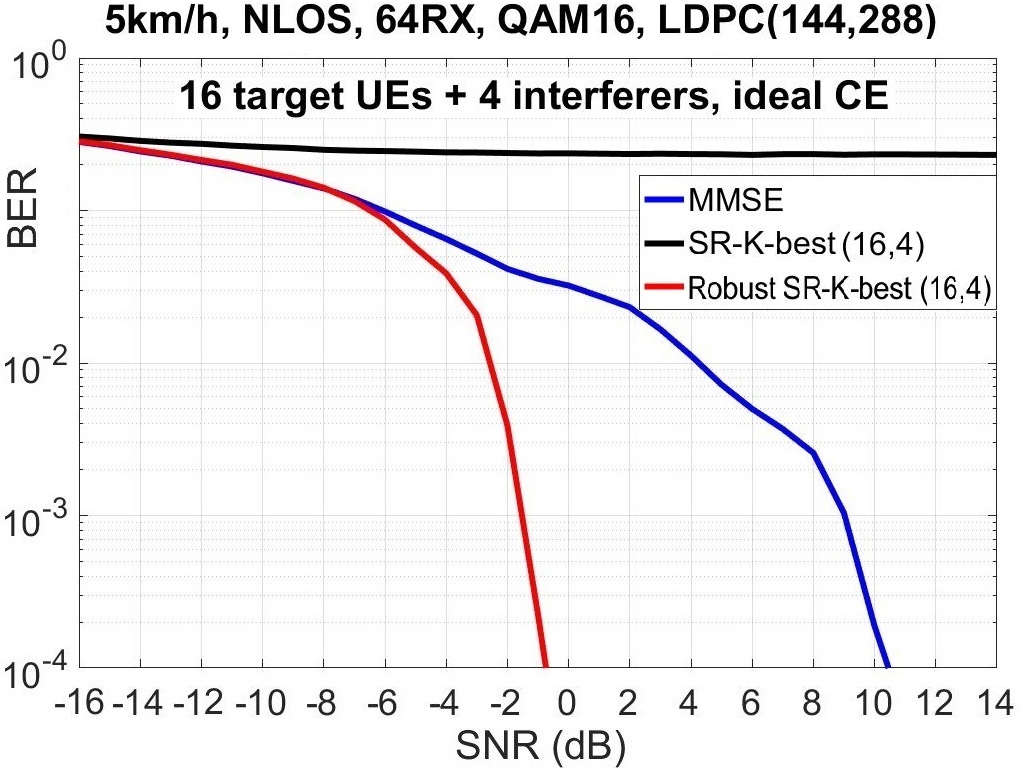}
\caption{
Performance in interference channel with ideal CE
}
\label{fig5}
\end{figure}

\begin{figure}[h]
\centering
\includegraphics[width=0.9\columnwidth]{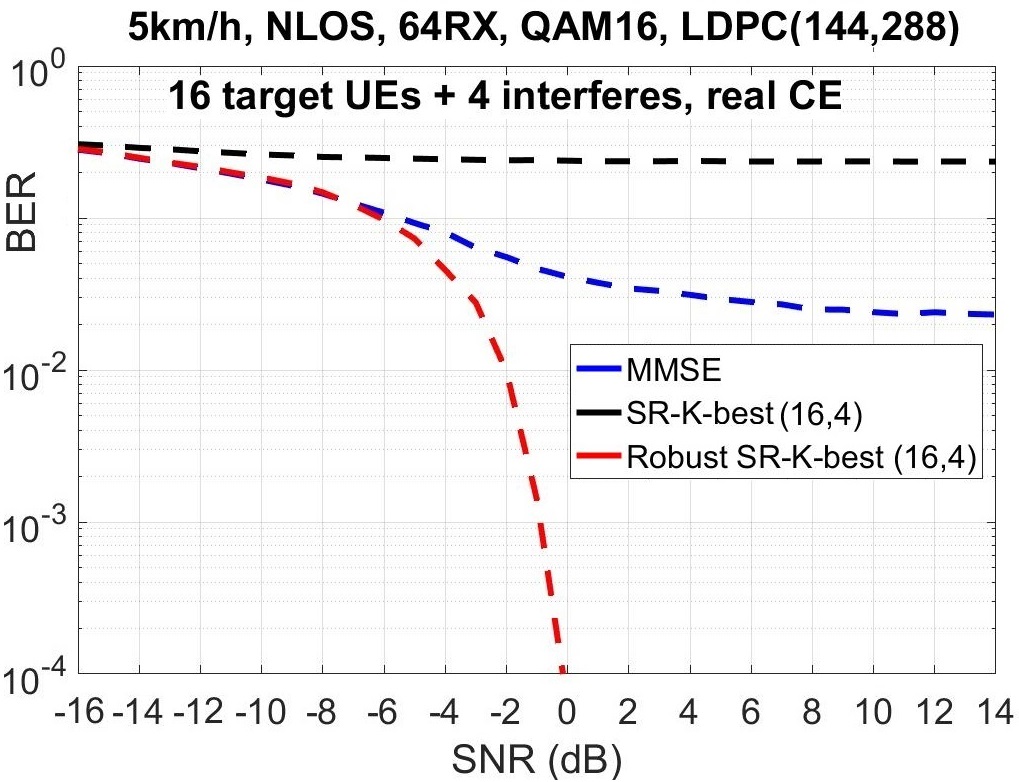}
\caption{
Performance in interference channel with real CE
}
\label{fig6}
\end{figure}

Simulation results of SR-K-best detector with parameters \eqref{all} are shown in \Fig{fig5} and \Fig{fig6} for $16$ target users with both ideal and realistic channel estimation in an interference environment with $4$ unknown users. It is clear that a common SR-K-best detector demonstrates the gain in AWGN channel only and does not detect users at the presence of interference, while the MMSE detector \eqref{GrindEQ__91_} is quite stable. It happens because the SR-K-best detector does not consider matrix ${\boldsymbol{R}}_{\boldsymbol{uu}}$. To overcome this problem, let us modify the SR-K-best pre- and post-processing algorithms. Equation \eqref{GrindEQ__1_} is given by:
\begin{equation}\label{eq1}
\boldsymbol{y}=\boldsymbol{H}\boldsymbol{x}+\boldsymbol{u},
\end{equation}
where $\boldsymbol{u}=\boldsymbol{i}+\boldsymbol{n}$ is the interference + noise signal. To avoid correlation between antennas in signal $\boldsymbol{u}$ we multiply \eqref{eq1} by $\sqrt {\boldsymbol{R^{-1}_{uu}}}$ as follows:
\begin{equation}
\sqrt {\boldsymbol{R^{-1}_{uu}}} \boldsymbol{y}=\sqrt {\boldsymbol{R^{-1}_{uu}}}\boldsymbol{H}\boldsymbol{x}+\sqrt {\boldsymbol{R^{-1}_{uu}}}\boldsymbol{u},
\end{equation}
\begin{equation}\label{eq2}
\boldsymbol{y_1}=\boldsymbol{H_1}\boldsymbol{x}+\boldsymbol{u_1},
\end{equation}
where $\boldsymbol{y_1}=\sqrt {\boldsymbol{R^{-1}_{uu}}} \boldsymbol{y}$, $\boldsymbol{H_1}=\sqrt {\boldsymbol{R^{-1}_{uu}}}\boldsymbol{H}$ and $\boldsymbol{u_1}=\sqrt {\boldsymbol{R^{-1}_{uu}}}\boldsymbol{u}$. Autocorrelation matrix of vector $\boldsymbol{u_1}$ can be calculated as:
\[
\begin{aligned}
\mathbb{E}(\boldsymbol{u_1}\boldsymbol{u_1^H}) & = \mathbb{E} \left(\sqrt{\boldsymbol{R^{-1}_{uu}}}\boldsymbol{u}\boldsymbol{u^H}\sqrt{\boldsymbol{R^{-1}_{uu}}} \right) \\
& = \sqrt{\boldsymbol{R^{-1}_{uu}}} \mathbb{E} \left( \boldsymbol{u}\boldsymbol{u^H} \right) \sqrt{\boldsymbol{R^{-1}_{uu}}} \\
& = \sqrt{\boldsymbol{R^{-1}_{uu}}} \boldsymbol{R_{uu}} \sqrt{\boldsymbol{R^{-1}_{uu}}} \\
& = I,
\end{aligned}
\]
therefore, signal $\boldsymbol{u_1}$ represents uncorrelated noise. After applying QR decomposition to matrix $\boldsymbol{H_1}$ we achieve the equation:
\begin{equation}\label{eq3}
\boldsymbol{y_1}=\boldsymbol{Q_1R_1}\boldsymbol{x}+\boldsymbol{u_1}
\end{equation}
The result of multiplying \eqref{eq3} by $\boldsymbol{Q_1^{H}}$ is given by:
\[\boldsymbol{Q_1^H}\boldsymbol{y_1}=\boldsymbol{Q_1^HQ_1R_1}\boldsymbol{x}+\boldsymbol{Q_1^Hu_1}\]
\begin{equation}\label{eq4}
\boldsymbol{y_2}=\boldsymbol{R_1}\boldsymbol{x}+\boldsymbol{u_2}
\end{equation}
where $\boldsymbol{y_2}=\boldsymbol{Q_1^H}\boldsymbol{y_1}$ is the modified input vector and $\boldsymbol{u_2}=\boldsymbol{Q_1^Hu_1}$ is the new white noise signal. In fact, the users sorting can be used with matrix $\boldsymbol{H_1}$ in \eqref{eq2} and further applying SR-K-best algorithm on \eqref{eq4} to achieve a fine detection performance. However, performance will be improved if we apply the MMSE detection to get $\boldsymbol{\widehat{x}}$ from \eqref{eq4}. Remember, that $\boldsymbol{u_2}$ is a white noise with the variance of $\sigma ^2 = 1$, therefore, linear MRC detector \eqref{GrindEQ__92_} can be utilized to \eqref{eq4} to calculate $\boldsymbol{\widehat{x}}$ as follows:
\begin{equation}
\boldsymbol{\widehat{x}}={ (\boldsymbol{I} + \boldsymbol{R_1^H} \boldsymbol{R_1}) }^{-1}{\boldsymbol{R_1^H} \boldsymbol{y_2}}. 
\end{equation}
\[\boldsymbol{R_1^H} \boldsymbol{y_2}=({\boldsymbol{I} + \boldsymbol{R_1^H} \boldsymbol{R_1} })\boldsymbol{\widehat x} +\boldsymbol{u_3},\]
\[\boldsymbol{y_2}= ( (\boldsymbol{R_1^H})^{-1}+\boldsymbol{R_1}) \boldsymbol{\widehat x} +(\boldsymbol{R_1^H})^{-1}\boldsymbol{u_3},\]
where $u_3$ is the leftover noise after MMSE. Finally, we achieve the equation:
\begin{equation}\label{eq5}
\boldsymbol{y_2}= \boldsymbol{H_2} \boldsymbol{\widehat x} + \boldsymbol{u_4},
\end{equation}
where $\boldsymbol{H_2}=((\boldsymbol{R_1^H})^{-1}+\boldsymbol{R_1})$ and noise $\boldsymbol{u_4}=(\boldsymbol{R_1^H})^{-1}\boldsymbol{u_3}$. Then we again apply QR decomposition to matrix $\boldsymbol{H_2}$ to achieve the following equation:
\begin{equation}\label{eq6}
\boldsymbol{y_2}= \boldsymbol{Q_2R_2}\boldsymbol{\widehat x} + \boldsymbol{u_4}
\end{equation}
The result of multiplying \eqref{eq6} by $\boldsymbol{Q_2^{H}}$ is given by:
\[\boldsymbol{Q_2^H}\boldsymbol{y_2}=\boldsymbol{Q_2^HQ_2R_2}\boldsymbol{\widehat x}+\boldsymbol{Q_2^Hu_4}\]
\begin{equation}\label{eq7}
\boldsymbol{y_3}= \boldsymbol{R_2}\boldsymbol{\widehat x} + \boldsymbol{u_5},
\end{equation}
where $\boldsymbol{y_3}=\boldsymbol{Q_2^H}\boldsymbol{y_2}$ and white noise is defined as $\boldsymbol{u_5}=\boldsymbol{Q_2^Hu_4}$. Finally, equations \eqref{eq5} and \eqref{eq7} are the best choice for SR-K-best detection according to out simulations and defines as Robust SR-K-best detector. Users sorting and permutation matrix $\boldsymbol{P}$ calculation should be used with matrix $\boldsymbol{H_2}$, while post-processing is implemented on the upper triangular matrix $\boldsymbol{R_2}$. Simulation results are presented in \Fig{fig5} and \Fig{fig6} for the Robust SR-K-best with parameters of \eqref{all} in the interference scenario. It should be noticed, that the developed nonlinear algorithm is robust to both interference and channel estimation errors (DFT-based channel estimation from \cite{A50} was implemented) and outperforms the linear MMSE. 

\section{Conclusion}

We proposed a new pre-processing for the nonlinear detector structure, which demonstrates significant performance gain compared to the MMSE detector in scenarios with unknown interference. Moreover, the achieved detector is robust to channel estimation errors. Simulation results show that the proposed algorithm is quite stable with non-ideal channel estimation in both AWGN and interference channels, while the MMSE detector demonstrates BER saturation after the LDPC decoder. Traditionally, the linear detector is known to be the best solution for unknown interference scenarios or non-ideal channel estimation, but our results say that even in these cases performance can be enhanced due to nonlinear detector nature. Simulation results with modulation QAM16 and LDPC(144,288) decoder are provided for the co-located scenario with $64$ antennas of Massive MIMO: $16$ single antenna target users and $4$ interferers in the $3D$-UMa model of 5G QuaDRiGa 2.0 channel. 


\end{document}